%% file: l.tex
\documentclass[aps,preprintnumbers,eqsecnum,amsmath,amssymb,nofootinbib]{revtex4}  
\usepackage{graphicx} 
\usepackage{bm}

\begin{document}
\title{Theoretical analysis of the leptonic decays $\bm{B\to \ell \ell \ell'\bar\nu_{\ell'}}$}
\author{Mikhail A.~Ivanov$^a$ and Dmitri Melikhov$^{a,b,c}$}
\affiliation{
$^a$Joint Institute for Nuclear Research, Dubna, 141980, Russia\\
$^b$D.~V.~Skobeltsyn Institute of Nuclear Physics, M.~V.~Lomonosov Moscow State University, 119991, Moscow, Russia\\
$^c$Faculty of Physics, University of Vienna, Boltzmanngasse 5, A-1090 Vienna, Austria}
\date{\today}
\begin{abstract} 
We discuss the amplitude of the $B\to l^+l^-l'\nu'$ decays and the differential decay rate $d^2\Gamma/dq^2dq'^2$,
$q$ the momentum of the $l^+l^-$ pair emitted from the electromagnetic vertex and $q'$ the momentum of the $l'\nu'$
pair emitted from the weak vertex. For the relevant form factors, we construct dispersion representations in $q^2$
which consistently take into account the Ward identity constraints at $q^2=0$ and the contributions of light 
vector resonances. This allows a consistent description of the form factors in the range $0 < q^2 \le 1$ GeV$^2$ 
that saturates around 99\% of the decay rate.
The differential decay rate behaves at small $q^2$ as $d\Gamma(B\to l^+l^-l' \nu')/dq^2\propto 1/q^2$ in the
limit $m'_l=0$, but contains also more singular contribution of order ${m'^2_l}/q^4$,
which we take into account.  
For the case $m_l'\le m_l$, the latter may be neglected and one obtains a mild logarithmic
dependence of $\Gamma(B\to l^{+}l^{-}l'\nu')$ on $m_l$.
For the case $m_l\ll m'_l$, however, the ${m'^2_l}/q^4$ terms dominate the decay rate leading to
$\Gamma(B\to l^{+}l^{-}l'\nu')\sim m'^2_l/m^2_l$. 
We find the following features of the four-lepton $B$-decays:  
(i) The decay rates $\Gamma\left(B\to \mu^+\mu^-(\mu \nu_\mu,e\nu_e)\right)$ are fully dominated 
by the region of light vector resonances $q^2\simeq M_\rho^2, M_\omega^2$;
(ii) The decay rate $\Gamma(B\to e^{+}e^{-}e \nu_e)$ receives comparable contributions from the region near 
$q^2\sim 4m_e^2$ and from the resonance region; 
(iii) 
One finds a strong enhancement of the decay rate
$\Gamma(B\to e^+e^- \mu \nu_\mu)\sim m_\mu^2/m_e^2$ which is dominated by the region $q^2\sim 4m_e^2$
due to the terms $O(m_\mu^2/q^4)$ in the differential distribution. 
\end{abstract}
\maketitle
\section{Introduction}
\label{Sec_introduction}
In this paper we revisit the amplitude $B\to\gamma^* l'\nu'$: we discuss constraints imposed by gauge invariance,
construct dispersion representations for the corresponding form factors,
and obtain predictions for the differential distributions in the $B$-meson decays 
into four leptons in the final state, $B\to l^+l^-l'\nu'$. The latter reactions 
are being studied experimentally \cite{exp1,exp2,exp3,exp4}, thus requiring a proper theoretical understanding 
of the $B$-meson form factors into two currents.
By now, there have been a few theoretical papers \cite{sehgal,nikitin,bharucha2021,beneke2}, where $B$-decays into two
lepton pairs have been studied.

\begin{figure}[h!]
\includegraphics[height=3.7cm]{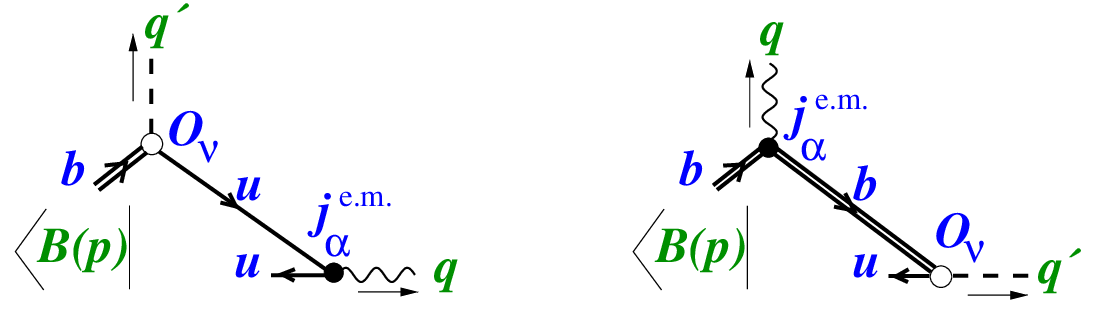}
\caption{\label{Fig:1} 
Feynman diagrams describing the amplitude (\ref{def}).}
\end{figure}

The $B\to \gamma^*l'\nu'$ amplitude (see Fig.~\ref{Fig:1}) may be parametrized via Lorenz-invariant
form factors as follows: 
\begin{eqnarray}
\label{def}
T_{\alpha\nu}(q,q'|p)=i\int dx\,  e^{i q x} \langle 0| T\{j^{\rm e.m.}_\alpha(x),\bar u(0){\cal O_\nu}b(0))\}|\bar B_u(p)\rangle = 
\sum_i L^{(i)}_{\alpha\nu}(q,q')F_i(q'^2,q^2)+\ldots, \quad p=q+q', 
\end{eqnarray}
with $q'$ the momentum of the weak $b\to u$ current, 
and $q$  the momentum of the electromagnetic current.
In Eq.~(\ref{def}), ${\cal O_\nu}=\gamma_\nu, \gamma_\nu\gamma_5$ and $j_\alpha^{\rm e.m.}$
is the conserved electromagnetic current 
\begin{eqnarray}
\label{jem}
j_\alpha^{\rm e.m.}(0)=e Q_b \bar b(0)\gamma_\alpha b(0) + e Q_u \bar u(0) \gamma_\alpha u(0).
\end{eqnarray} 
The quantities $L^{(i)}_{\alpha\nu}(q,q')$ represent the transverse Lorents 
structures, $q^\alpha L^{(i)}_{\alpha\nu}(q,q')=0$, and the dots stand for the longitudinal part which is constrained by the 
conservation of the electromagnetic current, $\partial_\alpha j_\alpha^{\rm e.m.}=0$,
and the equal-time commutation relations. 

The form factors $F_i(q'^2,q^2)$ are complicated functions of two variables, $q'^2$ and $q^2$;
the general properties of these objects in QCD have been studied recently in \cite{ims}.
Noteworthy, gauge invariance provides essential constraints on some of 
the form factors describing the transition of the $B$-meson into the real photon, i.e.,
at $q^2=0$ \cite{bmns2001,m2002,kruger,kmn2016}.

In the past, theoretical analyses focused on a family of similar reactions, namely, 
the $B\to\gamma l^+l^-$ and $B\to \gamma l\nu$ decays
(see, e.g., \cite{aliev,korchemsky,kou,mn2004,beneke1,kmn2018,beneke2018,ivanov,zwicky2019,bobeth,zwicky2021}); 
these processes are described by the same form factors as four-lepton $B$-decays,
but evaluated at a zero value of one of the momenta squared. 
The corresponding form factors depend on one variable $q'^2$, $q'$ the momentum of the weak current; for instance,
for radiative leptonic decays $B\to \gamma l'\nu'$, one needs the form factors $F_i(q'^2,q^2=0)$. 

The four-lepton decay of interest, $B\to l^+l^-l'\nu'$, requires the form factors $F_i(q'^2,q^2)$ for $0<q^2,q'^2<M_B^2$. 
The dependence of the form factors on the variable $q'^2$ can be predicted reasonably well: 
there are no hadron resonances in the full decay region $0<q'^2<M_B^2$, and the $q'^2$-dependence of the form factors 
is determined to a large extent by the influence of the beauty mesons with the appropriate quantum numbers; all these 
mesons are heavier than the $B$-meson and therefore lie beyond the physical decay region of the variable $q'^2$. 
The calculation of the $q^2$-dependence of the form factors is a much more difficult task: light vector mesons
$V=\rho^0,\omega,...$ lie 
in the physical decay region and should be properly taken into account. At $q^2$ in the region
of light vector meson resonances, the form factors cannot be obtained directly in pQCD \cite{ims}.
Here considerations based on the explicit account of these light 
vector resonances---including their finite width effects---are mandatory; the resonance contributions of interest 
may be unambiguously expressed via the weak $B\to V$ form factors.
Then, at $q^2=0$, gauge-invariance constrains the values of the form factors. 
These features allow us to calculate the form factors $F_i(q'^2,q^2)$ in the region $0<q^2\le 1-2$ GeV$^2$,
which dominates the four-meson decay rates and obtain consistent predictions for the latter.

Let us now turn to the differential distributions. 
After summing over the polarizations of the final leptons, the square of the amplitude of the $B\to l^+l^-l'\nu'$ decay may
be written in the following form: 
\begin{eqnarray}
\label{A2expansion}
|A|^2=|A|^2_0+|A|^2_{m_l'^2}+\dots,     
\end{eqnarray}
where $|A|^2_0$ corresponds to the massless leptons, $m_l=m_l'=0$, $|A|^2_{m_l'^2}$ is the term proportional to ${m_l'^2}$ which
provides the most singular behaviour of the amplitude, and the dots stand for those terms which yield 
negligible contributions to the differential and to the integrated decay rate compared to $|A|^2_0$ and may be safely omitted.
Among the terms given by the dots in (\ref{A2expansion}) one finds also the terms $O(m_l^2/q^4)$ [$q^4\equiv (q^2)^2$], but the
contribution of the latter both to the differential and to the integrated decay rates may be neglected.

Due to the gauge-invariance constraints on the form factors, one finds 
\begin{eqnarray}
\label{Gamma0}
|A|^2_0 \propto 1/q^2.      
\end{eqnarray} 
This property was already emphasized in \cite{beneke2} where it was pointed out that the naive behaviour $1/q^4$,
reported earlier in \cite{nikitin}, is unphysical. Nevertheless, we find it useful to present here an
explicit derivation of the constraints on the amplitude imposed by gauge invariance.  
The term $|A|^2_0$ yields the contribution to the integrated decay rate $\Gamma(B\to l^+l^-l'\nu')$
that has a mild logarithmic dependence $\propto\log(m_l^2)$. 

The term $|A|^2_{m_l'^2}$, for which we derive an explicit expression, is proportional to $m_l'^2$
but has a more singular behaviour at $q^2\to 0$ compared to $|A|^2_0$: 
\begin{eqnarray}
\label{Gamma2}
|A|^2_{m_l'^2} \propto m_l'^2/q^4.       
\end{eqnarray}
The $|A|^2_{m_l'^2}$ contribution to the differential decay rate is negligible compared to the contribution of $|A|^2_0$
in the full kinematical region of $B$-decay and may be safely omitted except for one case:
If $m_l\ll m_l'$, the contribution of
$|A|^2_{m_l'^2}$ dominates over $|A|^2_0$ in the vicinity of the end point $q^2=4m_l^2$.
Moreover, in this case $|A|^2_{m_l'^2}$ gives the dominant contribution to the decay rate
$\Gamma(B\to l^+l^-l'\nu')\propto m_l'^2/m_l^2$.

We shall demonstrate that these essential qualitative features of the $q^2$-distribution at small $q^2$ yield important
consequences for the theoretical estimates of the $B\to l^+l^-l'\nu'$ branching fractions:
\begin{itemize}
\item[(i)] The branching fraction ${\rm Br}(B\to \mu^+\mu^-(\mu\nu_\mu , e\nu_e)$ is dominated by the region of $q^2$ around the 
light vector resonances whereas the region of small $q^2$ yields a much smaller contribution; 
\item[(ii)] The branching fraction ${\rm Br}(B\to e^+e^-e\nu)$ receives comparable contributions from the resonance
region and the end-point region near $q^2=4m_e^2$.
\item[(iii)] The branching fraction ${\rm Br}(B\to e^+e^-\mu \nu_\mu)$ is fully dominated by the end-point region $q^2=4m_e^2$. 
\end{itemize}
Noteworthy, in all these cases the region $q^2>1$ GeV$^2$ contributes less than 1\% of the decay rate.

\section{Constraints on the transition form factors \label{sec_constraints}}
We now discuss the requirements imposed by the electromagnetic gauge invariance on the $B\to \gamma^*$
transition amplitudes
$\langle  \gamma^*(q)|\bar u \left\{ \gamma_\nu,\gamma_\nu\gamma_5\right\} b|\bar B_u(p)\rangle$ 
induced by the vector and the axial-vector charged currents.\footnote{Appendix A provides the relations between the
amplitudes containing $B_q$ and $\bar B_q$ mesons.}
The corresponding form factors are functions of two variables, $q^2$ and $q'^2$, 
where $q'$ is the momentum of the weak $b\to u$ current, 
and $q$ is the momentum of the electromagnetic current, and $p=q+q'$. 
Gauge invariance provides constraints on some of the form factors describing the transition 
of $B_u$ to the real photon, $q^2=0$. 

\subsection{Form factors of the vector weak current}
In case of the vector charged quark current $\bar u\gamma_\nu b$, 
the gauge-invariant amplitude contains one Lorentz structure and 
one dimensionless form factor $F_V(q'^2,q^2)$:
\begin{eqnarray}
\label{vector}
T_{\alpha,\nu}=i\int dx e^{i q x}
\langle 0| T\left\{ j^{\rm e.m.}_\alpha(x), \bar u\gamma_\nu b(0)\right\}|\bar B_u(p)\rangle=
e\,\epsilon_{\nu\alpha q'q} \frac{F_V(q'^2,q^2)}{M_B}. 
\end{eqnarray}
The amplitude is transverse, $q^\alpha T_{\alpha,\nu}=0$, and contains no contact term. 
It is free of the kinematical singularities so gauge invariance provides no constraints on $F_V(q'^2,q^2=0)$. 
The contribution of the vector charged quark current to the amplitude of the $B\to \gamma^* l'\nu'$ decay reads 
\begin{eqnarray}
A_{\rm vector}(B\to \gamma^* l'\nu')= e \frac{G_F}{\sqrt2}\,V_{ub}\,\bar l'\gamma_\nu (1-\gamma_5)\nu' \,\varepsilon_\alpha^*(q)
\epsilon_{\nu\alpha q'q} \frac{F_V(q'^2,q^2)}{M_B}. 
\end{eqnarray}
\subsection{Form factors of the axial-vector weak current}
For the axial-vector current,  $\bar u\gamma_\nu\gamma_5 b$, the corresponding amplitude is more complicated: 
it contains three independent gauge-invariant structures and three form factors, $f_{1A}(q'^2,q^2), f_{2A}(q'^2,q^2), f_{3A}(q'^2,q^2)$, 
and in addition it has the contact term which is fully determined by the conservation of the electromagnetic 
current, $\partial^\alpha j_\alpha^{e.m.}=0$: 
\begin{eqnarray}
\label{axial-vector}
T^5_{\alpha,\nu}&=&i\int dx e^{i q x}
\langle 0| T\left\{ j^{\rm e.m.}_\alpha(x), \bar u\gamma_\nu\gamma_5 b(0)\right\}|\bar B_u(p)\rangle\nonumber
\\
&=&
i e\,\left(g_{\alpha\nu}-\frac{q_\alpha q_{\nu}}{q^2}\right)\,q'q\, f_{1A}
+
i e\,\left(q'_{\alpha}-\frac{q'q}{q^2}q_\alpha\right)\bigg\{
p_{\nu} f_{2A}+ q_{\nu} f_{3A}\bigg\}
+i e \,Q_{B}f_{B}\frac{q_\alpha p_\nu}{q^2}. 
\end{eqnarray} 
Here $Q_B\equiv Q_{\bar B_u}=Q_b-Q_u$ is the electric charge of the $\bar B_u$ meson and $f_{B}>0$ is defined according to  
\begin{eqnarray}
\langle 0|\bar u\gamma_\nu\gamma_5 b|\bar B_u(p)\rangle =if_{B}p_\nu. 
\end{eqnarray}
The last term in (\ref{axial-vector}) is just the longitudinal contact term mentioned above.
Let us briefly recall the standard way this term is obtained (see \cite{lm2006} for details): 
We calculate $q_\alpha T^5_{\alpha,\nu}$, represent $q_\alpha e^{i q x}=-i \frac{\partial}{\partial x^\alpha}e^{i qx}$,
and perform parts integration moving the derivative to the $T$-product. Making use of the conservation of
the electromagnetic current $\partial^\alpha j^{e.m.}_\alpha=0$, the only nonzero 
contribution comes from the differentiation of the $\theta$-functions defining the $T$-product,
leading to the equal-time commutator. In the end, we obtain 
($\hat Q$ is the time-independent electric charge operator, $\hat Q=\int d^3x\, j_0(x^0,\vec x)$) 
\begin{eqnarray}
q_\alpha T^5_{\alpha,\nu}= - \langle 0|[\hat Q, \bar u\gamma_\nu \gamma_5 b(0)]|B(p)\rangle=i Q_B f_B p_\nu.
\end{eqnarray}
This relation does not determine the longitudinal Lorentz structure in the unique way: one can, e.g., choose this structure in the form 
$p_\alpha p_\nu/pq$ \cite{bmns2001,m2002} or in the form $q_\alpha p_\nu/q^2$ \cite{kmn2016}. However, only the latter form, which is implemented in (\ref{axial-vector}),  
corresponds to the longitudinal part in the form of a contact term.\footnote{By definition, a contact term is a quantity represented 
by $\delta$-function and its derivatives in configuration space; therefore $q_\alpha p_\nu/q^2$ is a contact term whereas 
$p_\alpha p_\nu/pq$ is not a contact term according to the standard definition. For further details we refer to
\cite{currentalgebra}.} 
Obviously, different choices of the longitudinal part lead to redefinitions of the form factors $f_{iA}$ in the transverse
part of the amplitude \cite{KhodjamirianWyler}. 
The choice of the longitudinal structure in the form of a contact term $q_\alpha p_\nu/q^2$ is suggested by the structure
of the quark 
electromagnetic vertex and is preferable with respect to the analytic properties of the form factors $f_{iA}$ \cite{m2002}.

The projectors in (\ref{axial-vector}) contain kinematical singularities at $q^2=0$. 
These singularities however should not be the singularities of the physical amplitude,
as the spectrum of physical states does not contain a massless vector particle in the $q^2$-channel;
recall that the absence of massless vector mesons is a fundamental feature of QCD.
Therefore, as the consequence of gauge invariance and the property of the spectrum of hadrons in QCD,
we obtain the following relations between the form factors at $q^2=0$:  
\begin{eqnarray}
\label{constraint1}
&&\left[f_{1A}(q'^2,q^2)+f_{3A}(q'^2,q^2)\right]_{q^2=0}=0,\\
\label{constraint2}
&&[q'q \,f_{2A}(q'^2,q^2)]_{q^2=0}=Q_B f_B. 
\end{eqnarray} 
To implement these constraints at $q^2=0$, we write down dispersion representations for the form factors
$f_{1A},f_{2A},f_{3A}$ in the 
variable $q^2$ with one subtraction 
and determine the subtraction terms to satisfy (\ref{constraint1}) and (\ref{constraint2}).
Such representations have the following form 
\begin{eqnarray}
\label{f1A}
f_{1A}(q'^2,q^2)&=&\xi(q'^2)+q^2\int \frac{ds}{\pi s(s-q^2)}\rho_{1A}(q'^2,s),\\
\label{f2A}
f_{2A}(q'^2,q^2)&=&\frac{2 Q_B f_B}{M_B^2-q'^2}+q^2\int \frac{ds}{\pi s(s-q^2)}\rho_{2A}(q'^2,s),\\
\label{f3A}
f_{3A}(q'^2,q^2)&=&-\xi(q'^2)+q^2\int \frac{ds}{\pi s(s-q^2)}\rho_{3A}(q'^2,s).
\end{eqnarray}
The form factor $\xi(q'^2)$ is related to the form factor of the $B\to \gamma l'\nu'$ transition,
and for the spectral densities $\rho_{Ai}(q'^2,s)$ we 
will construct phenomenological expressions based on the contributions of the light vector resonances $\rho_0$ and $\omega$. 

Next, we should add the Bremsstrahlung contribution (i.e., the photon emitted from the lepton $l'$ in the final state) 
that in the limit of a massless lepton $m_{l'}=0$ reads  
\begin{eqnarray}
\label{brems}
A_{\rm Brems}(B\to \gamma^* l'\nu')=
ie\, Q_l \frac{G_F}{\sqrt2}\,V_{ub}\,\bar l'\gamma_\nu (1-\gamma_5)\nu'\, \varepsilon^*_\alpha(q) f_B\left(-g_{\alpha\nu}\right),\qquad Q_l=Q_B, 
\end{eqnarray}
The axial part of weak-transition amplitude $B\to \gamma^* l'\nu'$ then takes the form  
\begin{eqnarray}
&&A_{\rm axial}(B\to \gamma^* l'\nu')\nonumber\\
&=&i e\frac{G_F}{\sqrt2}\,V_{ub}\,\bar l'\gamma_\nu (1-\gamma_5)\nu'\,\varepsilon_\alpha^*(q)
\left\{\left(g_{\alpha\nu}-\frac{q_\alpha q_{\nu}}{q^2}\right)\,q'q\, f_{1A}
+
\left(q'_{\alpha}-\frac{q'q}{q^2}q_\alpha\right)\bigg[
p_{\nu} f_{2A} + q_{\nu} f_{3A}\bigg]
+Q_{B}\,f_{B}\frac{q_\alpha p_\nu}{q^2}\right\}\nonumber
\\
&&+ie\, Q_B \frac{G_F}{\sqrt2}\,V_{ub}\,\bar l'\gamma_\nu (1-\gamma_5)\nu'\, \varepsilon^*_\alpha(q) f_B\left(-g_{\alpha\nu}\right),   
\end{eqnarray} 
the last term being the Bremsstrahlung contribution (\ref{brems}). 

The amplitude may be simplified by taking into account that $q^\alpha \varepsilon^*_\alpha(q)=0$, yielding
\begin{eqnarray}
A_{\rm axial}(B\to \gamma^* l'\nu')&=&i e \frac{G_F}{\sqrt2}V_{ub}\bar l'\gamma_\nu (1-\gamma_5)\nu' \varepsilon_\alpha^*(q)
\nonumber\\
&&\times\bigg\{
(g_{\alpha\nu}\,q'q-q'_\alpha q_\nu)\left[f_{1A}-\frac{Q_Bf_B}{q'q}\right]
+q'_\alpha q_\nu \left[f_{2A}+f_{3A}+f_{1A}-\frac{Q_Bf_B}{q'q}\right]
+q'_\alpha q'_\nu f_{2A}\bigg\}.\nonumber\\
\end{eqnarray}
Introducing dimensionless form factors $F_{1A,2A}$ and $F'_{2A}$ 
\begin{eqnarray}
\frac{F_{1A}(q'^2,q^2)}{M_B}&=&f_{1A}(q'^2,q^2)-\frac{Q_Bf_B}{q'q},\nonumber\\
\frac{F_{2A}(q'^2,q^2)}{M_B}&=&f_{2A}(q'^2,q^2)+f_{3A}(q'^2,q^2)+f_{1A}(q'^2,q^2)-\frac{Q_Bf_B}{q'q},\nonumber\\
\frac{F'_{2A}(q'^2,q^2)}{M_B}&=&f_{2A}(q'^2,q^2), 
\end{eqnarray}
we find the final expression for the contribution of the axial-vector part of the quark current,
$\bar u\gamma_\nu\gamma_5b$, to the amplitude 
\begin{eqnarray}
  \label{Aaxial}
A_{\rm axial}(B\to \gamma^* l'\nu')= i e \frac{G_F}{\sqrt2}\,V_{ub}\bar l'\gamma_\nu (1-\gamma_5)\nu' \varepsilon_\alpha^*(q)
\left\{(g_{\alpha\nu}q'q-q'_\alpha q_\nu)\frac{F_{1A}}{M_B}+q'_\alpha q_\nu \frac{F_{2A}}{M_B}
+q'_\alpha q'_\nu \frac{F'_{2A}}{M_B}\right\}. 
\end{eqnarray}
The contribution of the Lorentz structure $q'_\alpha q'_\nu$ in (\ref{Aaxial}) is proportional to $m_l'$
but generates the most singular, $\sim m'^2_l/(q^2)^2$, contribution to the differential decay rate, and, respectively,
the enhanced, $\sim m'^2_l/m^2_l$, contribution to the integrated decay rate, see Section \ref{Sect4}.
The Lorentz structure $q'_\alpha q'_\nu$ can be neglected in most of the cases, except for the case $m_l\ll m'_l$. 

Notice that as follows from Eqs.~(\ref{constraint1}) and  (\ref{constraint2}),
the form factors $F_{2A}$ and $F'_{2A}$
for the real photon in the final state satisfy the conditions
(see also \cite{Lattice2021}):
\begin{eqnarray}
\label{F2A}
F_{2A}(q'^2,q^2=0)&=&0, \\
\label{Fprime2A}
F'_{2A}(q'^2,q^2=0)&=&\frac{2Q_Bf_BM_B}{M_B^2-q'^2}. 
\end{eqnarray} 
The conditions (\ref{F2A}) are (\ref{Fprime2A}) are crucial as they determine the behavior of the differential distributions
in $B\to lll\nu$ at small $q^2$. For our parametrization of the amplitude in the form (\ref{axial-vector}),
the condition (\ref{F2A}) comes out as a direct consequence of Eqs.~(\ref{constraint1}) and (\ref{constraint2}).
Physics of course does not depend on the parametrization of the amplitude, but we find the parametrization
(\ref{axial-vector}) particularly convenient for the analysis of $B$-decays.  


\section{The $B\to \gamma l'\nu'$ transition}
We now illustrate the way the well-known formulas for the amplitude and the differential distribution
in the leptonic radiative $B$ decay, $B\to \gamma l'\nu'$, emerge. 

Since, as the consequence of gauge invariance, $F_{2A}(q'^2,q^2=0)=0$, only the form factors $F_{1A}$ and $F_V$
contribute to the amplitude for
the real photon and the {\it massless lepton} in the final state, and one finds the $B\to \gamma l'\nu'$ amplitude: 
\begin{eqnarray}
A(B\to \gamma l'\nu')=i e \frac{G_F}{\sqrt2}\,V_{ub}\bar l'\gamma_\nu (1-\gamma_5)\nu' \varepsilon_\alpha^*(q)
\bigg\{
(g_{\alpha\nu}q'q-q'_\alpha q_\nu)\frac{F_{A}(q'^2)}{M_B}+i \epsilon_{\nu\alpha q'q}\frac{F_{V}(q'^2)}{M_B}
\bigg\},
\end{eqnarray}
where $F_{A}(q'^2)\equiv F_{1A}(q'^2,q^2=0)$ and $F_{V}(q'^2)\equiv F_{V}(q'^2,q^2=0)$. 

The differential decay rate (for the massless lepton $m_{l'}=0$) takes a simple form:  
\begin{eqnarray}
  \frac{d\Gamma(B\to \gamma l'\nu')}{dE_\gamma}=
  \frac{G_F^2V_{ub}^2}{48\pi^2}M_B^4\alpha_{\rm e.m.}\,x_\gamma^3(1-x_\gamma)\big(|F_A|^2+|F_V|^2\big), 
\quad x_\gamma=2E_\gamma/M_B,
\end{eqnarray}
where $M_B E_\gamma=pq=q'q$, $E_\gamma$ the photon energy in the $B$-meson rest frame.


\section{\label{Sect4} The $B^-\to l^+l^-l'^-\bar \nu'$ transition}
The amplitude of the $B\to ll l'\nu'$ transition is readily obtained from the amplitudes of the 
$B\to \gamma^*l'\nu'$ transitions induced by the vector and the axial quark current by performing the replacement  
\begin{eqnarray}
\varepsilon^*_\alpha(q)\to - e \frac{\bar l\gamma_\alpha l}{q^2},
\end{eqnarray}
so we obtain
\begin{eqnarray}
A(B\to ll l'\nu')= i e^2 \frac{G_F}{\sqrt2}\,V_{ub}\cdot 
\bar l\gamma_\alpha l \cdot  \bar l'\gamma_\nu (1-\gamma_5)\nu' \frac{1}{q^2}
\bigg\{
(g_{\alpha\nu}q'q-q'_\alpha q_\nu)\frac{F_{1A}}{M_B}
+q'_\alpha q_\nu\frac{F_{2A}}{M_B}+q'_\alpha q'_\nu\frac{F'_{2A}}{M_B}
+ i \epsilon_{\nu\alpha q'q}\frac{F_{V}}{M_B}
\bigg\}.
\nonumber\\
\end{eqnarray}
By summing over the lepton polarizations and integrating over the phase space of the $l^+l^-$ 
pair and the $l'\nu'$ pair, one obtains the explicit analytic expression for the double differential distribution.
Neglecting term proportional to the lepton masses $m'_l$ and $m_l$, one obtains (see also \cite{beneke2}):\footnote{
Our form factors $f_{1A,2A,3A}$ are related to the form factors used in \cite{beneke2} as 
$f_{1A}=F_1/q'q$, $f_{2A}=(Q_B f_B-q^2 F_4)/q'q$, $f_{3A}=(-F_1 -q^2 F_6 +q^2 F_4)/q'q$. 
Our form factors $F_{1A,2A}$ are related to the form factors of \cite{beneke2} as 
$F_{A\perp}=\frac{q'q}{pq}F_{1A}$, $\tilde F_{A\parallel}=-F_{2A}+\frac{q^2}{pq}F_{1A}$,
and $F_{A\parallel}=-F_{2A}-F_{1A}\frac{4q^2q'^2}\lambda$.
}
\begin{eqnarray}
\label{A2}
|A|^2_0=&&\frac{G_F^2}{2}\,V_{ub}^2\,
\frac43 \frac83\frac{e^4}{(q^2)^2}\frac{1}{M_B^2}\nonumber\\
&&\times\bigg[2 F_{V}F^*_{V}\bar\lambda\, q^2q'^2+F_{1A}F^*_{1A}\left\{ 2 (q'q)^2+q^2q'^2\right\}q^2q'^2
  +(F_{1A} F_{2A}^*+F_{1A}^* F_{2A})\bar\lambda\, q^2q'^2+F_{2A}F_{2A}^*\bar\lambda^2\bigg],
\end{eqnarray}
with $\bar\lambda\equiv (q'q)^2-q^2q'^2=\frac{1}{4}\lambda(M_B^2,q^2,q'^2)$, where
$$\lambda(a,b,c)=(a-b-c)^2-4bc$$ is the triangle function.  
The factors $4/3$ and $8/3$ in Eq.~(\ref{A2}) result from the summation over polarizations of massless leptons 
coming from the electromagnetic vertex and from the weak vertex, respectively. 
Noteworthy, the expression in large square brackets behaves at small $q^2$ as $\propto q^2$, 
because of the constraint $F_2(q'^2,q^2=0)=0$.

More singular terms $\sim 1/q^4$ arise when one considers the effects of nonzero lepton masses.
Most of them can be safely neglected except for the contribution proportional to $m_l'^2$:
\begin{eqnarray}
\label{A2mlprime}
|A|^2_{m_l'^2}=&&\frac{G_F^2}{2}\,V_{ub}^2\,
\frac43\frac{e^4}{(q^2)^2}\frac{1}{M_B^2}\bigg[|F'_{2A}|^2  4m'^2_l (q'^2- 4m'^2_l)\bar\lambda\bigg]. 
\end{eqnarray}
This term is negligible compared to (\ref{A2}) in the full kinematical $B$-decay region
except for a close vicinity of the end point $q^2=4m_l^2$ in the case $m_l\ll m_l'$ . The 
term (\ref{A2mlprime}) leads to a singular $\sim m_l'^2/q^4$ contribution to the differential decay rate 
and the $\sim m'^2_l/m_l^2$ contribution to the integrated decay rate. The latter turns out 
to dominate the integrated decay rate in the case $m_l\ll m_l'$. 
Noteworthy, the relevant form factor at $q^2=0$ is fixed by the Ward identity
and contains only well-known parameters, Eq.~(\ref{Fprime2A}). Finally, we write  
\begin{eqnarray}
  |A|^2=|A|_0^2+|A|^2_{m'^2_l}+\dots,
\end{eqnarray}   
where the dots stand for those terms proportional to the lepton masses, $m_l$ and $m'_l$,
which may be safely neglected in the full kinematical region. 

The double differential distribution then takes the form (we display separately all numerical
factors according to the definition of the differential distribution):
\begin{eqnarray}
\label{Gamma}
\frac{d^2\Gamma(B\to ll l'\nu')}{dq^2dq'^2}=
\frac{(2\pi)^4}{2M_B}\frac{1}{(2\pi)^{12}}
\frac{\pi\lambda^{1/2}(q^2,m_l^2,m_l^2)}{2q^2}
\frac{\pi\lambda^{1/2}(q'^2,m_{l'}^2,0)}{2q'^2}
\frac{\pi\lambda^{1/2}(M_B^2,q^2,q'^2)}{2M_B^2}|A|^2.  
\end{eqnarray}
The kinematical constrains on the variables $q^2$ and $q'^2$ come from the $\lambda$-functions in Eq.~(\ref{A2}) and read 
\begin{eqnarray}
4m_l^2\le q^2, \quad  m_{l'}^2 \le q'^2,\quad \sqrt{q^2}+\sqrt{q'^2}\le M_B.
\end{eqnarray} 
First, let us notice that because of the gauge-invariance constraint (\ref{F2A}), 
one finds the behaviour $|A|_0^2\propto 1/q^2$ and not $\propto 1/(q^2)^2$ as may be naively obtained when the 
gauge-invariance constraint is not taken into account.
Such terms in $|A|^2$ lead to a mild logarithmic dependence of the integrated decay rate on $m_l$. 
Second, there are terms $\propto m_l'^2/(q^2)^2$ in $|A|^2$, which emerge from $|A|_{m_l'^2}^2$;
these terms are however essential only in a specific case $m_l\ll m_l'$. They lead to the low-$q^2$
enhancement of the decay rate as $m_l'^2/m_l^2$.  

We emphasize that the double differential distribution $d^2\Gamma(B\to l^+l^-l'\nu')/dq^2 dq'^2$
is easily calculable due to the fact that the leptons emitted from the electromagnetic vertex and the
lepton emitted from the weak vertex have different flavours; no exchange diagrams emerge in this case and one 
obtains the explicit analytic expression for the double differential distribution.

\section{The form factors and the differential distributions}
\subsection{Modelling the form factors}
The form factors $F_{1A,2A}(q'^2,q^2)$ are obtained from the form factors $f_{1A,2A,3A}$, using for the latter  
the $q^2$-dispersion representation with one subtraction at $q^2=0$,
Eqs.~ (\ref{f1A}), (\ref{f2A}), and (\ref{f3A}). The subtraction procedure allows us to incorporate
the constraints imposed by gauge invariance. 

Similarly, for the form factor $F_V(q'^2,q^2)$ a single-subtracted dispersion representation in $q^2$ is used:  
The form factors $F_A(0,q'^2)$ and $F_V(0,q'^2)$ should be equal to each other at the leading order of the
double $1/E_\gamma$ ($2 M_B E_\gamma=M_B^2-q'^2$) and $1/M_B$ expansions in QCD \cite{korchemsky}.
To satisfy this requirement, we make a subtraction in
$F_V(q^2,q'^2)$ at $q^2=0$ and add a subtraction term $F_V(q'^2)$.  

Furthermore, we assume that the spectral densities are saturated by the light vector-meson
resonances $\rho^0$ and $\omega$ in the $q^2$-channel 
and, since these resonances emerge in the physical region of the $B$-decay of interest, we take into account the finite-width 
effects of these resonances.
In the end, we come to the following expressions 
\begin{eqnarray}
\label{FA1}
F_{1A}(q'^2,q^2)&=&F_{A}(q'^2)-\frac{Q_B f_B M_B}{q'q}-
q^2
\sum\limits_{V=\rho^0,\omega}\bigg(
\frac{1}{M_V^2}\frac{2M_B(M_B+M_V)}{M_B^2-M_V^2-q'^2}\frac{M_V f_V}{M_V^2-q^2-i\Gamma_V(q^2)M_V}A_1^{B\to V}(q'^2)
\bigg),\nonumber\\
\\
\label{FA2}
F_{2A}(q'^2,q^2)&=&-q^2 M_B
\sum\limits_{V=\rho^0,\omega}
\frac{1}{M_V^2}\frac{2 M_V f_V}{M_V^2-q^2-i\Gamma_V(q^2)M_V}
\left[
\frac{M_B+M_V}{M_B^2-M_V^2-q'^2}A_1^{B\to V}(q'^2)-\frac{A_2^{B\to V}(q'^2)}{(M_B+M_V)}
\right]
\nonumber\\
&&+Q_Bf_B\left(\frac{2M_B}{M_B^2-q'^2}-\frac{2M_B}{M_B^2-q'^2-q^2}\right),
\\
\label{FV}
F_V(q'^2,q^2)&=&F_V(q'^2)-q^2M_B
\sum\limits_{V=\rho^0,\omega}\bigg(
\frac{1}{M_V^2}
\frac{M_V f_V}{M_V^2-q^2-i\Gamma_V(q^2)M_V}\frac{2 V^{B\to V}(q'^2)}{M_B+M_V}\bigg). 
\end{eqnarray}

Let us discuss the expressions above: 

\noindent 
$\bullet$ 
The form factors $F_A(q'^2)$ and $F_V(q'^2)$ describe the $B\to \gamma l'\nu'$ transition; 
they emerge as subtraction terms at $q^2=0$ in the $q^2$-disperison representations for the form factors 
$F_{1A,V}(q'^2,q^2)$. The form factors $F_A(q'^2)$ and $F_V(q'^2)$ 
are equal to each other at the leading order of the double $1/E_\gamma$ ($2 M_B E_\gamma=M_B^2-q'^2$) and 
$1/M_B$ expansions in QCD \cite{korchemsky} but differ in the subleading orders \cite{mn2004,beneke1,beneke2018}:  
\begin{eqnarray}
\label{leet1}
F_{A}(q'^2)&=&-\frac{Q_u f_B M_B}{2 E_\gamma\lambda_B}+\frac{Q_b f_B M_B}{2 E_\gamma m_b}+O(Q_u f_B M_B/E_\gamma^2),\\
\label{leet2}
F_{V}(q'^2)&=&-\frac{Q_u f_B M_B}{2 E_\gamma\lambda_B}-\frac{Q_b f_B M_B}{2 E_\gamma m_b}+O(Q_u f_B M_B/E_\gamma^2)
\end{eqnarray}
The magnitude of the form factors $F_A(q'^2)$ and $F_V(q'^2)$ is determined to a large extent by the
parameter $\lambda_B$,
the inverse moment of the $B$-meson light-cone distribution amplitude $\phi_B$ \cite{korchemsky}. 
The value of $\lambda_B$ presently has a large uncertainty: for instance,
\cite{beneke1} makes use of $\lambda_B(1\,{\rm GeV})=0.35$ GeV; 
the sum-rule estimate of \cite{kou} led to $\lambda_B(1\,{\rm GeV})=0.57$ GeV; 
Ref.~\cite{BraunIvanovKorchemsky2004} obtained $\lambda_B(1\,{\rm GeV})=0.46\pm 0.11$ GeV; 
a recent NLO analysis of \cite{zwicky2021} reported $\lambda_B(1\,{\rm GeV})=0.36\pm0.11$ GeV;
the results of calculating $F_A(q'^2)$ and $F_V(q'^2)$ \cite{mn2004} using the
dispersion approach \cite{m} correspond to a relatively 
large value $\lambda_B(1\,{\rm GeV})=0.657$ GeV.
Obviously, the uncertainty in the parameter $\lambda_B$ dominates the uncertainty
in the differential distributions $d\Gamma(B\to lll'\nu')/dq^2$ at small $q^2$.

In \cite{kmn2018}, the form factors $F_A(q'^2)$ and $F_V(q'^2)$ have been calculated in a broad range
$0<q'^2<25$ GeV$^2$ using the dispersion approach of \cite{m}. It was found that the monopole form 
(\ref{leet1}) and (\ref{leet2}) describes the results of our calculation for $0< q'^2 <15$ GeV$^2$ with a few \%
accuracy, whereas at $q'^2\sim 25$ GeV$^2$ the monopole formula overestimates the calculated form factors
by $\sim$ 20\%. Nevertheless, taking into account a large uncertainty in the present knowledge of the
parameter $\lambda_B$, we find it eligible to use the monopole form (\ref{leet1}) and (\ref{leet2})in the full
kinematically allowed region of $q'^2$ and consider the variation of $\lambda_B$ in the range
$\lambda_B(1\,{\rm GeV})=(0.5\pm 0.15)$ GeV. 

\vspace{.2cm}
\noindent  
$\bullet$
In the region $0.4 \le q^2({\rm GeV}^2) \le 0.9$,
where light vector meson resonances show up in the differential distributions,
the form factors of interest cannot be calculated using perturbative QCD \cite{ims}.
To calculate the form factors $F_{1A,2A,V}(q'^2,q^2)$ 
in this region of $q^2$ and for any $q'^2$ appropriate for the four-lepton decay, we make use of the dispersion representations 
and assume \cite{kmn2018,nikitin} that they may be saturated by the intermediate $\rho^0$ and $\omega$-states 
in the $q^2$-channel.\footnote{We would like to notice that no relative phase
between the $\rho^0$ and $\omega$ contributions to the form factors 
$F_i(q'^2,q^2)$ as proposed in \cite{nikitin} may emerge: these form factors contain a sum over the intermediate
states $|V\rangle\langle V|$, so 
even if one introduces arbitrary complex phases in the states $|V \rangle$, these phases appear both in the decay constants 
$f_V$ and the $B\to V$ weak form factors such that they finally drop out from $F_i(q'^2,q^2)$.}
Since the light neutral vector mesons lie in the physical decay region of $q^2$, it is necessary to take into account their finite 
$q^2$-dependent width $\Gamma_V(q^2)$. For a relatively broad 
$\rho$-meson the function $\Gamma_V(q^2)$ takes into account the effects of the $2\pi$ intermediate states; the appropriate formulas 
are given in \cite{nachtmann}. In practical calculations, we use a simplified expression 
which takes into account the correct threshold behaviour of the 
$\rho\to \pi\pi$ phase space: $\Gamma_{\rho^0}(q^2)=\theta(q^2-4m_\pi^2)(1-4m_\pi^2/q^2)^{3/2}/(1-4m_\pi^2/M_\rho^2)^{3/2}\Gamma_{\rho^0}$. 
For a narrow $\omega$-meson, we take an approximation of constant width. (This approximation is not fully theoretically clean:
the imaginary part of the propagator of the vector meson should vanish below the threshold in the corresponding decay channel.
This means that $\Gamma_V(q^2)$ should vanish below the corresponding light-meson threshold.
But for a narrow $\omega$-meson the effect is tiny). Table~\ref{Table:fV} gives the meson parameters 
entering the form factors Eqs.~(\ref{FA1}), (\ref{FA2}), and (\ref{FV}). 

Notice that $\Gamma_V(q^2)$ takes into account the contribution of continuum of light pseudoscalar mesons; in this way
we effectively take into account the contribution of hadron continuum to the spectral densities of the form
factors $F_{1A,2A,V}(q'^2,q^2)$ \cite{nachtmann}. In the end, one finds that the nonresonance $q^2$-region, $q^2\ge 1.0$ GeV$^2$
gives a small contribution to the
decay width of the $B\to lll'\nu'$ decay. This agrees with the expectations of the analysis of \cite{beneke2}. 
 
\begin{table}[t!]
\caption{\label{Table:fV}
Meson parameters entering the expressions for the form factors, Eqs.~(\ref{FA1}), (\ref{FA2}), and (\ref{FV}). 
Data from \cite{ball2007,pdg}.}
\centering
\begin{tabular}{|c|c|c|c|c|}
\hline
 $f_B$       & $\sqrt2 f_\rho^0$   & $3\sqrt2 f_\omega$  & $\Gamma_{\rho^0}$  &   $\Gamma_{\omega}$     \\
\hline
190 MeV      &  216\ MeV         &       190 MeV      & 150 MeV          &   8.49 MeV            \\
\hline
\end{tabular}
\end{table}

\vspace{.2cm}
\noindent  
$\bullet$
The contribution of the light vector mesons $V=\rho^0,\omega$ to the form factors $F_{1A,2A,V}(q'^2,q^2)$
is {\it unambiguous} (cf. \cite{beneke2}) and are expressed via the form factors
$A_1^{B\to V}(q'^2)$, $A_2^{B\to V}(q'^2)$, and $V^{B\to V}(q'^2)$ 
describing the weak decay $B\to V$. In spite of many efforts to calculate these form factors in a
broad kinematical decay region $0 < q'^2 < M_B^2$, our knowledge of these quantities is not very accurate.
Table \ref{table:weak_form_factors} presents some selected results for the 
Relevant form factors: although the central values of the form factors at $q'^2=0$ from different
approaches are in reasonable agreement 
with each other, the uncertainties vary from an ``educated guess'' of 10\% for \cite{ms2000} to
almost 50\% in \cite{gubernari2019}. 
The uncertainties in these form factors, along with the uncertainty in the parameter $\lambda_B$,
is the second main source of the uncertainty 
in the theoretical predictions for $B\to l^+l^-l'\nu'$ decays. Appendix \ref{AppendixB} summarizes the necessary
parametrizations of the form factors used in our numerical estimates. 

\begin{table}[t!]
\caption{\label{table:weak_form_factors}
Selected theoretical predictions for the weak form factors describing $B$ decays into light vector mesons.
The form factors from \cite{ms2000,ballzwicky2005,ivanov2} are expected to have a 10-15\% uncertainty. 
To obtain the form factors for $B^-\to \rho^0$ and $B^-\to \omega$ decays,  
the numbers given in this Table should be multiplied by the isotopic factor $1/\sqrt{2}$.}
\centering
\begin{tabular}{|c|c|c|c|c|c|c|}
\hline
    Ref.                 & $A_1^{B\to \rho}(0)$  & $ A_1^{B\to \omega}(0)$ & $A_2^{B\to \rho}(0)$ & $A_2^{B\to \omega}(0) $ & $V^{B\to \rho}(0)$ & $V^{B\to \omega}(0)$  \\
\hline
\cite{ms2000}            &  0.26              &     $-$   &    0.24      &  $-$    &   0.31   &  $-$  \\
\cite{ballzwicky2005}    &  0.24              &    0.22   &    0.22      &  0.20   &   0.32   &  0.29 \\
\cite{ivanov2}           &  0.26              &     $-$   &    0.24      &  $-$    &   0.28   &  $-$  \\
\cite{gubernari2019}     &  $0.22\pm 0.1$     &     $-$   &$0.19\pm 0.11$&  $-$    &$0.27\pm 0.14$ &  $-$  \\
\hline
\end{tabular}
\end{table}
\subsection{The differential distributions}
\begin{figure}[h!]
\centering
\begin{tabular}{cc}
\includegraphics[width=8cm]{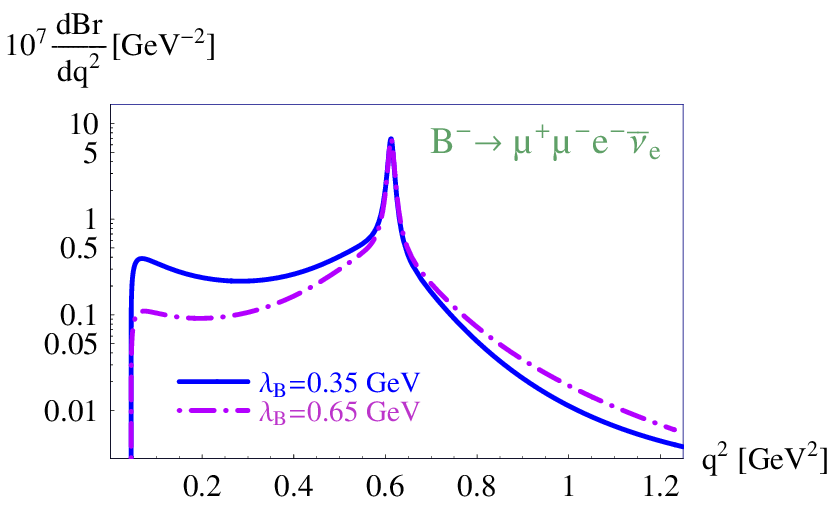}\qquad  &  \qquad \includegraphics[width=8cm]{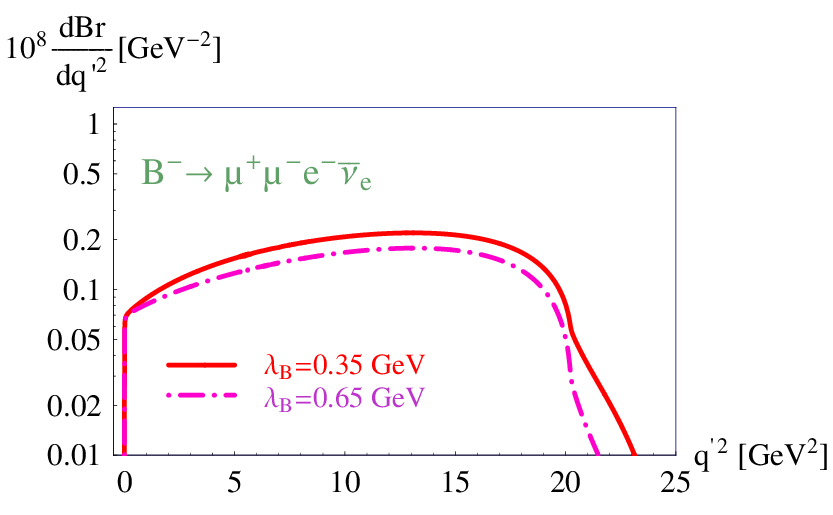}\\
(a) & (b)\\
\end{tabular}
\caption{\label{Plot:1} 
The differential distributions $d\Gamma(B\to\mu^+\mu^-e \nu_e)/dq^2$ (a) and $d\Gamma(B\to\mu^+\mu^-e \nu_e)/dq'^2$ (b), 
for the weak transition $B\to V$ form factors from \cite{ms2000}. 
Solid lines corresponds to $\lambda_B=0.35$ GeV, dashed lines correspond to $\lambda_B=0.65$ GeV.}
\end{figure}

With the analytic expressions for the form factors, Eqs. (\ref{FA1}), (\ref{FA2}), and (\ref{FV}) at hand, Eqs.~(\ref{Gamma}) 
give the differential distributions in $B\to\mu^+\mu^- e \nu_e$ decays, Fig.~\ref{Plot:1}. Here we use 
$V_{ub}=0.004$ and $\tau_{B^-}=1.638\,10^{-12}$ s. The Plots show the impact of the parameter $\lambda_B$ on the differential 
distributions $d\Gamma/dq^2$ and $d\Gamma/dq'^2$. 

\begin{figure}[h!]
  \begin{center}
  \includegraphics[width=7cm]{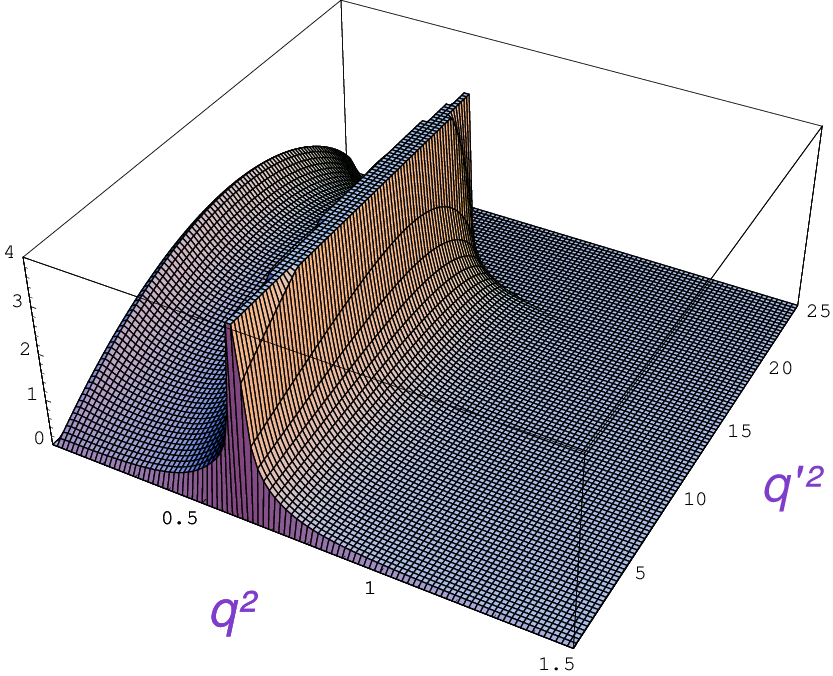}
  \end{center}
  \caption{\label{Plot:2} The double differential distribution $d^2\Gamma(B\to\mu^+\mu^- e \nu_e)/dq^2dq'^2$ calculated for
$\lambda_B=0.65$ GeV and the form factors from \cite{ms2000}.}
\end{figure}

Figure \ref{Plot:2} shows the double differential distributions calculated for $\lambda_B=0.65$ GeV
and the form factors from \cite{ms2000}.

\section{Discussion and Conclusions}
Our results are summarized below: 


\vspace{.1cm}
\noindent
1. Gauge invariance provides essential constraints on the amplitude of Eq.~(\ref{axial-vector}): 
\begin{eqnarray}
\label{C0}
T^5_{\alpha,\nu}&=&i\int dx e^{i q x}
\langle 0| T\left\{ j^{\rm e.m.}_\alpha(x), \bar u\gamma_\nu\gamma_5 b(0)\right\}|\bar B_u(p)\rangle. 
\end{eqnarray}
We emphasize that for a consistent analysis of the $B\to l^+l^-l'\nu'$ amplitude it is necessary to start with the
amplitude 
(\ref{C0}) and properly parametrize this amplitude taking into account all constraints imposed by the electromagnetic
gauge invariance and analyticity. Taking into account these constraints leads to 
\begin{eqnarray}
\label{C1}
F_{2A}(q'^2,q^2=0)=0,\quad F'_{2A}(q'^2,q^2=0)=\frac{2Q_B f_B M_B}{M_B^2-q'^2}. \nonumber
\end{eqnarray}
These relations determine the behaviour of the differential decay rate  
$d^2\Gamma(B\to lll'\nu')/dq^2dq'^2$ at small $q^2$ (see Eqs.~(\ref{A2}) and (\ref{A2mlprime})).

\vspace{.1cm}
\noindent
2. For the form factors $F_{V,1A,2A}(q'^2,q^2)$, describing the amplitude of $B\to lll'\nu'$ decay,
we obtained dispersion representations in $q^2$ with one subtraction.
This allows us to take into account properly both the constraints imposed by gauge invariance
at small $q^2$ and the contributions of vector mesons ($V$); the latter involve the weak
form factors describing $B\to V$ decays. 
Assuming that the light vector mesons $\rho^0$ and $\omega$ saturate the
spectral densities, we obtained analytic representations for the form factors in a
broad range of $q^2$ and $q'^2$.

Our assumption may seem oversimplified in comparison with a sophisticated analysis
of the form factors presented in \cite{beneke2}. Moreover, our spectral representations saturated by
merely light vector mesons do not reproduce the correct $q^2$-behaviour and overshoot the form factors
at large values of $q^2$, where the form factors may be calculated using OPE \cite{ims,beneke2};
this means that our form factors do not produce realistic differential distributions at large values of $q^2$. 

Nevertheless, our approach has a certain advantage compared to that of \cite{beneke2}: Making use of the
once-subtracted $q^2$-dispersion representations allows us to take properly into account both the
gauge-invariance constraints 
and the resonance contributions to the form factors: the latter may be calculated unambiguously
and are found to be nozero, see (\ref{F2A}). On the other hand, in Ref.~\cite{beneke2} 
the resonance contributions to $F_{2A}$ are omitted in order to satisfy the
gauge-invariance constraints.

A proper description of the resonance region of $q^2$ is crucial as it produces
the bulk of the $B\to \mu^+\mu^- l\nu$ cross-section and nearly a half of the $B\to e^+e^- l \nu$ cross section
(the other half comes from the region of small $q^2$). So, from the point
of view of obtaining numerical predictions, we find it eligible to trade the proper description of the region of
$0< q^2< 1$ GeV$^2$ against overestimating the contribution of the region of large $q^2$ which anyway,
even with our overshot form factors, contributes at less than a percent level.

\vspace{.1cm}
\noindent
3. We derived an explicit analytic expression for the differential distributions $d\Gamma/dq^2 dq'^2$ in $B\to lll'\nu'$
decays including the $O(m_l'^2/q^4)$ terms which provide the most singular behaviour of the differential distribution
at small $q^2$. We then obtained numerical predictions for the differential and the 
integrated branching ratios of the $B\to l^+l^-l'\nu'$ decays.

To illustrate the lepton-mass effects, Table \ref{Table:BRs} presents the numerical
results for various decay modes. For the modes with identical particles in the final state,
instead of the full decay rate that includes the exchange diagrams, Table \ref{Table:BRs} shows the quantity 
$\Gamma(B\to lll'\nu')|_{m_{l'}=m_l}$. The full results for the identical leptons in the final state are
discussed in the next item. 
\begin{table}[h!]
\caption{\label{Table:BRs}
The branching ratios of the $B\to l^+l^- l\nu$ integrated over the specific $q^2$
ranges for the form factors from \cite{ms2000} and $\lambda_B=0.5$ GeV.  
For $B\to e^+e^- e\nu_e$ and $B\to \mu^+\mu^- \mu\nu_\mu$ the results for
$\Gamma(B\to lll'\nu')|_{m_{l'}=m_l}$ are given.}
\centering
\begin{tabular}{|c|c|c|c|c|c|}
\hline
Mode    & $q^2=[4m_e^2,4m_\mu^2]$ & $q^2=[4m_\mu^2, 0.4$ GeV$^2]$& $q^2=[0.4$ GeV$^2$, 1 GeV$^2$] & $q^2$=[1 GeV$^2$, $q^2_{\rm max}]$ & Total \\
\hline
$e^+e^- \mu\nu_\mu$& $6.05\cdot 10^{-7}$ & $6.72\cdot 10^{-9}$ & $2.51\cdot 10^{-8}$ & $4.14\cdot 10^{-10}$ &  $6.38 \cdot 10^{-7}$  \\
$\mu^+\mu^- e\nu_e$& $-$               & $5.42\cdot 10^{-9}$ & $2.42\cdot 10^{-8}$ & $4.10\cdot 10^{-10}$ &  $3.01 \cdot 10^{-8}$  \\     
$\mu^+\mu^-\mu\nu_\mu$& $-$            & $5.41\cdot 10^{-9}$ & $2.41\cdot 10^{-8}$ & $4.07\cdot 10^{-10}$ &  $3.00 \cdot 10^{-8}$ \\
$e^+e^-e\nu_e$  & $1.96\cdot 10^{-8}$   & $6.81\cdot 10^{-9}$  & $2.52 \cdot 10^{-8}$ & $4.17\cdot 10^{-10}$ & $5.21 \cdot 10^{-8}$ \\
\hline
\end{tabular}
\end{table}

\vspace{.2cm}
\noindent
4. We now present the numerical results including the estimated uncertainties.
The uncertainties in our predictions come from the two main sources:

\noindent 
(i) The uncertainty in the parameter $\lambda_B$, which governs the behaviour of the $q^2$-differential distributions
at small $q^2\le 0.4$ GeV$^2$ but has an impact on the $q'^2$-distributions in the broad range of $q'^2$. 
We allow the parameter $\lambda_B$ to vary in the range $\lambda_B=0.35-0.65$ GeV (the lower values of this range
has been advocated in several analyses \cite{korchemsky,beneke1,zwicky2021} whereas the upper values
of $\lambda_B$ is obtained in explicit model calculations \cite{kou,mn2004}).

\noindent
(ii) The uncertainties in the $B\to \omega, \rho$ weak form factors $V,A_1,A_2$,
which mainly govern the differential distributions in the region
of $q^2=(0.4-0.9)$ GeV$^2$. To obtain the numerical estimates, we use as the basic scenario the form factors
calculated in \cite{ms2000}, and in order to estimate the uncertainties allow a
15\% uncertainty on these form factors. 
We take into account a 10\% suppression of the $B\to \omega$ form factors compared to the corresponding $B\to \rho$ 
form factors according to \cite{ballzwicky2005}.

Taking into account these uncertainties, we obtain the following estimates 
\begin{eqnarray}
  {\rm Br}(B\to\mu^+\mu^- e\nu_e)&=&(3.01\;{^{+0.53}_{-0.19}}|_{\lambda_b}\pm 0.82|_{\rm\, weak\, ffs})\, 10^{-8},\\ 
{\rm Br}(B\to e^+e^- \mu\nu_\mu)&=&(6.38\;{^{+0.31}_{-0.12}}|_{\lambda_b}\pm 0.08|_{\rm\, weak\, ffs})\, 10^{-7}.
\end{eqnarray} 
We emphasize that the full integrated rate ${\rm Br}(B\to e^+e^-\mu\nu_\mu)$ is an order of magnitude larger than
${\rm Br}(B\to \mu^+\mu^-e\nu_e)$. The former is fully dominated by the region $4m_e^2 < q^2 < 4m_\mu^2$
where the distribution contains an enhancement factor $(f_B/M_B)^2(m_\mu/m_e)^2$ due to the $O(m_\mu^2)$ terms
in the amplitude (\ref{Gamma2}).  

For the decay $B\to l^+ l^- l^+ \nu_l$ ($l=\mu, e$) with identical positive-charged leptons in the final state,
the amplitude is given by the sum of direct and exchange diagrams, $A=A_{\rm dir}+A_{\rm exchange}$, 
and the phase space includes a factor 1/2 because of the presence of the 
identical particles in the final state. The phase-space integration of both $|A_{\rm dir}|^2$ and $|A_{\rm exchange}|^2$
leads to the same result, $\Gamma(B\to lll'\nu')|_{m_{l'}=m_l}$, and one can write (see, e.g., \cite{beneke2}):  
\begin{eqnarray}
\label{br_identical}
\Gamma(B\to lll\nu)=\Gamma(B\to lll'\nu')|_{m_{l'}=m_l}+\Gamma_{\rm interference}(B\to lll\nu), 
\end{eqnarray}
where $\Gamma_{\rm interference}(B\to lll\nu)$ is the phase-space integral of
$(A_{\rm dir} A^*_{\rm exchange}+A^*_{\rm dir} A_{\rm exchange})$. The interference term should be calculated numerically
as the integral over the phase space. A simple analytic result similar to Eqs.~(\ref{A2}) 
and (\ref{Gamma}) cannot be obtained. 
We have performed a numerical calculation of the branching fraction (\ref{br_identical}) and found that the interference
branching fraction leads to a very mild increase of the integrated branching fraction ${\rm Br}(B\to lll'\nu')$ at the
level of less than 1\% (our detailed results for the differential distributions for this case will be presented in \cite{im2022}). 
We report 
\begin{eqnarray}
\label{4mu}
{\rm Br}(B^+\to\mu^+\mu^- \mu^+\bar\nu_\mu)=
(3.02\;{^{+0.45}_{-0.25}}|_{\lambda_b}\pm 0.62|_{\rm\, weak\, ffs})\, 10^{-8}. 
\end{eqnarray}
This estimate agrees with the result of \cite{beneke2} and is only marginally
compatible with the upper limits obtained by the LHCb Collaboration
\cite{exp4} ${\rm Br}(B^+\to\mu^+\mu^- \mu^+\bar\nu_\mu)\le 1.6 \cdot \,10^{-8}$.
Recall, however, that the experimental upper bound applies certain kinematical cuts
whereas our result corresponds to the branching fraction integrated over the full
allowed region of the lepton momenta. 

For electrons in the final state, we find 
\begin{eqnarray}
\label{4e}
{\rm Br}(B^+\to e^+e^-e^+\bar\nu_e)=
(5.26\;{^{+2.60}_{-1.05}}|_{\lambda_b}\pm 0.70|_{\rm\, weak\, ffs})\, 10^{-8}.
\end{eqnarray} 

\acknowledgments
We are grateful to M.~Beneke and R.~Zwicky for valuable comments. 
D.~M.~gratefully acknowledges support from the Russian Foundation of
Basic Research under joint RFBR-CNRS grant 19-52-15022.

\input{l_Appendix.tex}
\input{l_Parametrizations.tex}

\input{l_References.tex}
\end{document}

%% file: l_Appendix.tex
\appendix
\section{Relations between the $B_q$ and $\bar B_q$ amplitudes}
Here we derive the relations between the amplitudes of $B_q$ and $\bar B_q$ mesons.
Such relations are obtained by applying charge conjugation.

The fermion field transforms under charge conjugation $\hat C$
($\hat C^2=1$, $\hat C^{-1}=\hat C$) as follows \cite{IZ}: 
\begin{eqnarray}
\label{Ctransform}
\hat C \psi \hat C&=&\eta_c{ \cal C} \bar \psi^T,\\
\hat C \bar \psi \hat C&=&\eta^*_c \psi^T ({\cal C}^T)^{-1}, 
\end{eqnarray}  
$|\eta_c|=1$, where the charge-conjugation matrix ${\cal C}$ is defined by the relation  
\begin{eqnarray}
{\cal C} \gamma_\mu^T  {\cal C}^{-1}&=&-\gamma_\mu 
\end{eqnarray}
and has the following properties: ${\cal C}^T=-{\cal C}$, ${\cal C}^{-1}=-{\cal C}$, ${\cal C}^2=-1$. 
In the Dirac representation of the $\gamma$-matrices, one can choose 
${\cal C}=i\gamma_0\gamma_2$ leading to 
\begin{eqnarray}
   {\cal C} \gamma_5^T  {\cal C}&=&-\gamma_5,\\
   {\cal C} (\gamma_\mu\gamma_5)^T {\cal C}&=&-\gamma_\mu\gamma_5,\\
   {\cal C} (\sigma_{\mu\nu})^T {\cal C}&=&\sigma_{\mu\nu},\\
   {\cal C} (\sigma_{\mu\nu}\gamma_5)^T  {\cal C}&=&\sigma_{\mu\nu}\gamma_5.
\end{eqnarray}
Making use of these relations, one obtaines the following expression for charge conjugation of bilinear currents
(of anticommuting) fermion operators:
\begin{eqnarray}
 \hat C (\bar \psi_1{\cal O}\psi_2)\hat C&=& - \bar \psi_2 ({\cal C}{\cal O}^T {\cal C})\psi_1,
\end{eqnarray}
leading to 
\begin{eqnarray}
\hat C (\bar \psi_1\psi_2)\hat C&=& \bar \psi_2\psi_1,\\
\hat C (\bar \psi_1\gamma_5 \psi_2)\hat C&=& - \bar \psi_2\gamma_5\psi_1,\\  
  \hat C (\bar \psi_1\gamma_\mu \psi_2)\hat C&=& - \bar \psi_2\gamma_\mu \psi_1,\\
  \hat C (\bar \psi_1\gamma_\mu\gamma_5 \psi_2) \hat C&=& \bar \psi_2\gamma_\mu\gamma_5 \psi_1,\\
  \hat C (\bar \psi_1\sigma_{\mu\nu} \psi_2) \hat C&=&-\bar \psi_2 \sigma_{\mu\nu}\psi_1,\\
  \hat  C (\bar \psi_1 \sigma_{\mu\nu}\gamma_5 \psi_2) \hat C&=&-\bar \psi_2\sigma_{\mu\nu}\gamma_5\psi_1.
\end{eqnarray}
The $C$-conjugate states are related to each other as follows (no arbitrary phase is implied): 
\begin{eqnarray}
  \hat C|B_q(p)\rangle=|\bar B_q(p)\rangle,
\end{eqnarray}
The QCD vacuum state is $C$-invariant, $\hat C|0\rangle=|0\rangle$. So, if we are going to consider QCD
effects in the amplitudes, we can apply $C$-conjugation and relate to each other the amplitudes 
\begin{eqnarray}
&&\langle 0|\bar q\gamma_\mu\gamma_5 b|\bar B_q(p)\rangle=i \bar f_B p_\mu,\\  
&&\langle 0|\bar b\gamma_\mu\gamma_5 q|B_q(p)\rangle=i f_B p_\mu
\end{eqnarray}
and obtain the relation $f_B= \bar f_B$.  
Similar relations may be obtained for more complicated amplitudes such as  
\begin{eqnarray}
\bar T^{{\cal O}}_{\alpha}(p,q)\equiv  i\int dx e^{iqx}\langle 0|T\{j_\alpha^{\rm e.m.}(x),\bar q(0){\cal O} b(0)\}|\bar B_q(p)\rangle,
  \label{T1}
  \\
  \label{T2}
T^{{\cal O}}_{\alpha}(p,q)\equiv i\int dx e^{iqx}\langle 0|T\{j_\alpha^{\rm e.m.}(x),\bar b(0){\cal O} q(0)\}|B_q(p)\rangle.
\end{eqnarray}
The parametrizations of these amplitudes are given via the same form factors but with appropriate sign
adjustments between (\ref{T1}) and (\ref{T2}). For instance, for the relation between the amplitudes
(we use in this Appendix the notation $\bar T$ for the amplitudes containing $\bar B_q$-meson in the initial state)
\begin{eqnarray}
\bar T_{\alpha\mu}&=&
i\int dx e^{iqx}\langle 0|T\{j_\alpha^{\rm e.m.}(x),\bar q(0)\gamma_{\mu} b(0)\}|\bar B_q(p)\rangle,\\
\bar T^5_{\alpha\mu}&=&
i\int dx e^{iqx}\langle 0|T\{j_\alpha^{\rm e.m.}(x),\bar q(0)\gamma_{\mu}\gamma_5 b(0)\}|\bar B_q(p)\rangle,\\
\bar T_{\alpha,\mu\nu}&=&
i\int dx e^{iqx}\langle 0|T\{j_\alpha^{\rm e.m.}(x),\bar q(0)\sigma_{\mu\nu} b(0)\}|\bar B_q(p)\rangle,\\
\bar T^5_{\alpha,\mu\nu}&=&
i\int dx e^{iqx}\langle 0|T\{j_\alpha^{\rm e.m.}(x),\bar q(0)\sigma_{\mu\nu}\gamma_5 b(0)\}|\bar B_q(p)\rangle,
\end{eqnarray}
and the corresponding amplitudes $T_{\alpha\mu}, T^5_{\alpha\mu}, T_{\alpha,\mu\nu}$, and $T^5_{\alpha,\mu\nu}$ as defined according to
(\ref{T1}) and (\ref{T2}), we obtain 
\begin{eqnarray}
\bar T_{\alpha\mu}(p,q)&=&T_{\alpha\mu}(p,q),\\
\bar T^5_{\alpha\mu}(p,q)&=&-T_{\alpha\mu}(p,q),\\ 
\bar T_{\alpha\mu\nu}(p,q)&=&T_{\alpha\mu\nu}(p,q),\\
\bar T^5_{\alpha\mu\nu}(p,q)&=&T^5_{\alpha\mu\nu}(p,q). 
\end{eqnarray}
In conclusion, the amplitudes (\ref{T1}) and (\ref{T2}) are related to each other by charge conjugation. 

%% file: l_Parametrizations.tex
\section{Parametrizations of the form factors 
\label{AppendixB}}
\subsection{$F_{V,A}(q'^2)$}

In our numerical estimates we use the following parametrizations for the form factors ($Q_u=2/3$, $Q_b=-1/3$): 
\begin{eqnarray}
F_{V}(q'^2)&=&-Q_u\frac{ M_B^2}{M_B^2-q'^2}\frac{f_B}{\lambda_B}- Q_b \frac{ M_B^2}{M_B^2-q'^2}\frac{f_B}{m_b},\\
F_{A}(q'^2)&=&-Q_u\frac{ M_B^2}{M_B^2-q'^2}\frac{f_B}{\lambda_B}+ Q_b \frac{ M_B^2}{M_B^2-q'^2}\frac{f_B}{m_b},
\end{eqnarray}
with $f_B=190$ MeV and $m_b=5$ GeV. The parameter $\lambda_B$ varies in the range $\lambda_B$=0.35-0.65 GeV.

\subsection{$V(q'^2), A_1(q'^2), A_2(q'^2)$}
All the form factors are parametrized as follows
\begin{eqnarray}
F_i(q'^2)=\frac{F_i(0)}{(1-\sigma^{(i)}_0 r)(1-\sigma_1^{(i)} r+\sigma^{(i)}_2 r^2)}, \qquad r\equiv q'^2/M_R^2. 
\end{eqnarray}
For the basic scenario of Melikhov, Stech \cite{ms2000} the parameters are given below and $M_R=M_{B*}=5.32$ GeV. 
{\it Notice}: all tables give $F_i(0)$ for the $B\to \rho^+$ transition. For $B\to \rho^0$ transition
$F_i(0)$ should be multiplied by isotopic factor $1/\sqrt{2}$. 
\begin{center}
\begin{tabular}{|c|c|c|c|}
\hline
                & $V^{B\to \rho^+}(0)$  & $A_1^{B\to \rho^+}(0)$  &  $A_2^{B\to \rho^+}(0)$   \\
\hline
$F(0)$          &  0.31             &     0.26     &    0.24    \\
$\sigma_0$      &   1               &    0         &    0     \\
$\sigma_1$      &  0.59             &    0.73      &    1.4   \\
$\sigma_2$      &  0                &    0.10      &    0.50   \\
\hline
\end{tabular}
\end{center}
$\bullet$ For $B\to \omega$ transition a reduction of the form factors at zero by 10\% compared to $B\to \rho^0$ was applied
following the estimates of \cite{ballzwicky2005}. The $q'^2$-dependence is taken the same as for $B\to \rho$. 

\noindent
$\bullet$
To estimate the uncertainty in the predictions for the rates, the range of the form factors from \cite{gubernari2019} was used
(see Table \ref{table:weak_form_factors}). 

\subsection{Resonance $q^2$-dependent width}

\begin{center}
\begin{tabular}{|c|c|c|c|c|}
\hline
 $f_B$       & $\sqrt2 f_\rho^0$   & $3\sqrt2 f_\omega$ & $\Gamma_{\rho^0}$  &   $\Gamma_{\omega}$     \\
\hline
190 MeV      &  216\ MeV         &       190 MeV      & 150 MeV          &   8.49 MeV            \\
\hline
\end{tabular}
\end{center}

For a relatively broad 
$\rho^0$-meson the function $\Gamma_V(q^2)$ takes into account the effects of the $2\pi$ intermediate states; the appropriate formulas 
are given in \cite{nachtmann}. In practical calculations, we use a simplified expression 
which takes into account the correct threshold behaviour of the 
$\rho\to \pi\pi$ phase space:
\begin{eqnarray}
\Gamma_{\rho^0}(q^2)=\theta(q^2-4m_\pi^2)(1-4m_\pi^2/q^2)^{3/2}/(1-4m_\pi^2/M_\rho^2)^{3/2}\Gamma_{\rho^0}.
\end{eqnarray}
For a narrow $\omega$-meson, we take an approximation of constant width.

%% file: l.bbl
\begin{thebibliography}{100}
\bibitem{exp1}
LHCb Collaboration (R.~Aaij et al.), 
{\it Search for the rare decay $B_s^0\to \mu^+\mu^-\mu^+\mu^-$}, 
Phys.~Rev.~Lett.~ {\bf 110}, 211801 (2013).
\bibitem{exp2}
ATLAS Collaboration (M.~Aaboud et al.),
{\it Study of the rare decays of $B^0_s$ and $B_0$ into muon pairs
from data collected during the LHC Run 1 with the ATLAS detector}, 
Eur.~Phys.~J.~{\bf C76}, 513 (2016).
\bibitem{exp3}
LHCb Collaboration (R.~Aaij et al.), 
{\it Search for decays of neutral beauty mesons into four muons}, 
JHEP {\bf 1703}, 001 (2017).
\bibitem{exp4}
LHCb Collaboration (R.~Aaij et al.), 
{\it Search for the rare decay $B^+\to \mu^+\mu^-\mu^+\nu_\mu$}, 
Eur.~Phys.~J.~{\bf C79}, 675 (2019).
\bibitem{sehgal}
Y.~Dincer and L.~M.~Sehgal, 
{\it Electroweak effects in the double Dalitz decay $B(s)\to l^+ l^- l'^+ l'^-$}, 
Phys.~Lett.~{\bf B556}, 169 (2003).
\bibitem{nikitin}
A.~V.~Danilina and N.~V.~Nikitin, 
{\it Four-Leptonic Decays of Charged and Neutral $B$ Mesons within the Standard Model}, 
Phys.~Atom.~Nucl.~{\bf  81}, 347 (2018), Yad.~Fiz. {\bf 81}, 331 (2018);
A.~Danilina N.~Nikitin, and K.~Toms, 
{\it Decays of charged $B$-mesons into three charged leptons and a neutrino}, 
Phys.~Rev.~{\bf D101}, 096007 (2020).
\bibitem{bharucha2021}
A.~Bharucha, B.~Kindra and N.~Mahajan, {\it Probing the structure of the $B$ meson with
$B\to lll'\nu'$}, ArXiv:2102.03193.
\bibitem{beneke2}
M.~Beneke, P.~B\"oer, P.~Rigatos, and K.~K.~Vos, 
{\it QCD factorization of the four-lepton decay $B\to lll\nu$},
Eur.~Phys.~J.~{\bf C81}, 638 (2021). 
\bibitem{ims}
M.~A.~Ivanov, D.~Melikhov, and S.~Simula, 
{\it Form factors for $B\to j_1 j_2$ decays into two currents in QCD}, 
Phys.~Rev.~{\bf D101}, 094022 (2020). 

\bibitem{bmns2001}
M.~Beyer, D.~Melikhov, N.~Nikitin, and B.~Stech, 
{\it Weak annihilation in the rare radiative $B\to \rho\gamma$ decay}, 
Phys.~Rev.~{\bf D64}, 094006 (2001). 

\bibitem{m2002}
D.~Melikhov, 
{\it Dispersion approach to quark binding effects in weak
decays of heavy mesons},  
Eur.~Phys.~Journal direct {\bf 4}, 2 (2002) [hep-ph/0110087]. 

\bibitem{kruger} 
F.~Kruger and D.~Melikhov,
{\it Gauge invariance and form-factors for the decay $B\to \gamma l^+ l^-$}, 
Phys.~Rev.~{\bf D67}, 034002 (2003).%

\bibitem{kmn2016}
A.~Kozachuk, D.~Melikhov, and N.~Nikitin, 
{\it Annihilation type rare radiative $B_{(s)}\to V\gamma$ decays}, 
Phys.~Rev.~{\bf D93}, 014015 (2016). 

\bibitem{aliev}
T.~M.~Aliev, A.~Ozpineci, and M.~Savci, 
{\it $B_q\to l^+l^-\gamma$ decays in light cone QCD}, 
Phys.~Rev.~{\bf D55}, 7059 (1997).

\bibitem{korchemsky}
G. Korchemsky, D. Pirjol, and T.-M. Yan,
{\it Radiative leptonic decays of $B$ mesons in QCD}, 
Phys.~Rev.~{\bf D61}, 114510 (2000).

\bibitem{kou}
P.~Ball and E.~Kou, 
{\it $B\to \gamma e\nu$ transitions from QCD sum rules on the light cone}, 
JHEP {\bf 0304}, 029 (2003). 

\bibitem{mn2004}
D. Melikhov and N. Nikitin, 
{\it Rare radiative leptonic decays $B_{(d, s)} \to \gamma l^+l^-$}, 
Phys.~Rev.~{\bf D70}, 114028 (2004).

\bibitem{beneke1}
M.~Beneke and J.~Rohrwild,
{\it B meson distribution amplitude from $B\to \gamma l \nu$}, 
Eur.~Phys.~J.~{\bf C71}, 1818 (2011).

\bibitem{kmn2018}
A. Kozachuk, D. Melikhov, and N. Nikitin, 
{\it Rare FCNC radiative leptonic $B_{s,d}\to \gamma l^+l^-$ decays in the Standard Model}, 
Phys.~Rev.~{\bf D97}, 053007 (2018).

\bibitem{beneke2018}
M.~Beneke, V.~M.~Braun, Y.~Ji, and Y.-B.~Wei, 
{\it Radiative leptonic decay $B\to \gamma \ell \nu_\ell$
with subleading power corrections}, 
JHEP {\bf 1807}, 154 (2018). 
\bibitem{ivanov}
S. Dubnicka, A. Z. Dubnickova, M. A. Ivanov, A. Liptaj,
P. Santorelli, and C. T. Tran, 
{\it Study of $B_s\to l^+l^-\gamma$ decays in covariant quark model}, 
Phys.~Rev.~{\bf D99}, 014042 (2019).

\bibitem{zwicky2019}
J.~Albrecht, E.~Stamou, R.~Ziegler, and R.~Zwicky, 
{\it Probing flavoured Axions in the Tail of $B_q\to\mu^+\mu^-$}, 
arXiv:1911.05018.

\bibitem{bobeth}
M.~Beneke, C.~Bobeth, and Y.-M.~Wang, 
{\it $B_{d,s}\to\gamma l^+l^-$ decay with an energetic photon}, 
JHEP {\bf 2012}, 148 (2020).

\bibitem{zwicky2021}
T.~Janowski, B.~Pullin, and R.~Zwicky, 
{\it Charged and neutral $\bar B_{u,d,s}\to \gamma$ form factors
from light cone sum rules at NLO}, arXiv:2106.13616. 


\bibitem{lm2006}
W.~Lucha and D.~Melikhov,
{\it Quark-hadron duality and hadron properties from correlators of pseudoscalar and axial currents}, 
Phys.~Rev.~{\bf D73}, 054009 (2006);
{\it OPE and sum rules for correlators of pseudoscalar and axial currents}, 
Phys.~Atom.~Nucl.~{\bf 70}, 891 (2007).

\bibitem{currentalgebra}
S.~Treiman, R.~Jackiw, D.~Gross, {\it Lectures on current algebra and its applications}, 
Princeton University Press, Princeton, New Jersey, 1972.  

\bibitem{KhodjamirianWyler}
A.~Khodjamirian and D.~Wyler, 
{\it Counting contact terms in $B\to V\gamma$ decays}, hep-ph/0111249.

\bibitem{Lattice2021}
A.~Desiderio {\it et al},  
{\it First lattice calculation of radiative leptonic decay rates of pseudoscalar mesons}, 
Phys.~Rev.~{\bf D103}, 014502 (2021).

\bibitem{BraunIvanovKorchemsky2004}
V.~M.~Braun, D.~Yu.~Ivanov, G.~P.~Korchemsky,
{\it The B meson distribution amplitude in QCD}, 
Phys.~Rev.~{\bf D69}, 034014 (2004).
\bibitem{m}
D.~Melikhov, 
{\it Form factors of meson decays in the relativistic constituent quark model},  
Phys.~Rev.~{\bf D53}, 2460 (1996);  
{\it Heavy quark expansion and universal form-factors in the quark model}, 
Phys.~Rev.~{\bf D56}, 7089 (1997).  

\bibitem{nachtmann}
D.~Melikhov, O.~Nachtmann, V.~Nikonov, and T. Paulus, 	
{\it Masses and couplings of vector mesons from the pion electromagnetic, 
weak, and $\pi\gamma$ transition form-factors}, 
Eur.~Phys.~J.~{\bf C34}, 345 (2004).

\bibitem{ball2007}
P.~Ball, G.~W.~Jones, and R.~Zwicky,
{\it $B \to  V \gamma$ beyond QCD factorisation}, Phys.~Rev.~{\bf D75}, 054004 (2007).

\bibitem{pdg}
Particle Data Group, C.~Patrignani {\it et al.}, Chin.~Phys.~{\bf C40}, 100001 (2016).

\bibitem{ms2000}
D.~Melikhov and B.~Stech,
{\it Weak form-factors for heavy meson decays: an update}, 
Phys.~Rev.~{\bf D62}, 014006 (2000).

\bibitem{ballzwicky2005}
P.~Ball and R.~Zwicky, 
{\it $B_{d,s}\to \rho,\omega, K^*,\phi$ decay form factors from light-cone sum rules reexamined},
Phys.~Rev.~{\bf D71}, 014029 (2005). 

\bibitem{ivanov2} 
M.~A.~Ivanov, J.~G.~Korner, S.~G.~Kovalenko, P.~Santorelli, and G.~G.~Saidullaeva, 
{\it Form factors for semileptonic, nonleptonic and rare $B(B_s)$ meson decays}, 
Phys.~Rev.~{\bf D85}, 034004 (2012). 

\bibitem{gubernari2019}
N.~Gubernari, A.~Kokulu and D.~van~Dyk, 
{\it $B\to P$ and $B\to V$ form factors from $B$-meson
light-cone sum rules beyond leading twist}, JHEP {\bf 1901}, 150 (2019). 

\bibitem{im2022} M.~A.~Ivanov and D.~Melikhov, {\it Theoretical analysis of $B\to l^+l^-l^+\nu_l$ decays 
with identical leptons in the final state}, in preparation. 

\bibitem{IZ}
  C.~Itzykson and J.-B.~Zuber, {\it Quantum Field Theory},
   McGraw-Hill Inc, 1980.

\end{thebibliography}
